\documentclass[
 twocolumn,
 english,superscriptaddress,unsortedaddress,showpacs,showkeys,secnumarabic,nobibnotes,final,prl
 ]
 {revtex4}

\usepackage{graphicx}
\usepackage{amssymb}
\usepackage{color}

\begin{document}

\title{Spread of Matter over a Neutron-Star Surface During Disk Accretion:\\
Deceleration of Rapid Rotation }

\author{N.A. Inogamov}

 \affiliation{Landau Institute for Theoretical Physics, RAS, Chernogolovka, 142432, Russian Federation}

 \affiliation{Max-Planck Institut fuer Astrophysik, Garching, D-85741, Germany}

 \author{R.A. Sunyaev}

 \affiliation{Max-Planck Institut fuer Astrophysik, Garching, D-85741, Germany}

 \affiliation{Space Research Institute, RAS, Moscow, 117997, Russian Federation}

\begin{abstract}
 The problem of disk accretion onto the surface of a neutron star with a weak magnetic field
   at a luminosity exceeding several percent of Eddington is reduced to the problem
     of the braking of a hypersonic flow with a velocity that is 0.4-0.5 of the speed of light
       above the base of the spreading layer -- a dense atmosphere made up of previously fallen matter.
 We show that turbulent braking in the Prandtl-Karman model with universally accepted coefficients
   for terrestrial conditions and laboratory experiments
     and a ladder of interacting gravity waves in a stratified quasi-exponential atmosphere
       at standard Richardson numbers lead to a spin-up of the massive zone that extends to the ocean
         made up of a plasma with degenerate electrons.
 Turbulent braking in the ocean at the boundary with the outer solid crust
   reduces the rotation velocity to the solid-body rotation velocity of the star.
 This situation should lead to strong heating of deep atmospheric layers and to the switch-off
   of the explosive helium burning mechanism.
 Obviously, a more efficient mechanism for the dissipation of a fast azimuthal flow in the atmosphere
   should operate in X-ray bursters.
 We show that a giant solitary gravity wave in the atmosphere can lead to energy dissipation
   and to a sharp decrease in azimuthal velocity in fairly rarefied atmospheric layers
     above the zone of explosive helium burning nuclear reactions.
 We discuss the reasons why this wave, that has no direct analog in the Earth's atmosphere or ocean, appears
   and its stability.
 We pose the question as to whether neutron stars with massive atmospheres,
   spun up to high velocities by accreting matter from a disk,
     can exist among the observed Galactic X-ray sources.
\end{abstract}





\maketitle

 \noindent Full text of the paper is published in Astronomy Letters, vol. 36, p. 848-894 (2010).

 \section{INTRODUCTION}

 The problem of a boundary layer under disk accretion onto a neutron star (NS)
   with a weak $(<10^8$G) magnetic field arose from the necessity of interpreting the observations
     of X-ray sources in low-mass binaries with neutron stars.
 In particular, we are talking about X-ray bursters called so because of recurrent X-ray bursts
   (Lewin et al. 1993).
 These bursts result from thermonuclear explosions in the matter that fell onto the NS after a preceding burst.
 The very interesting physical problem of a flow in a boundary layer has no analogs
   in the Earth's atmosphere and is inaccessible to modeling under laboratory conditions.
 It is reduced to the problem of inflow from a disk, braking, settling, and redistribution
   of the accreting matter along the meridian on the NS surface.
 On the NS, this matter forms a powerful flow that rotates azimuthally with a hypersonic velocity
   relative to the surface (the accretion rate is $10^{16}-10^{18}$ g s$\!^{-1}).$
 The flow is in the shape of an equatorial belt that moves with a velocity slightly less than the Keplerian one
   $(\sim 40-50\%$ of the speed of light) above a compact star with a radius of $\approx 10$ km.
 In this case, the acceleration due to gravity $g_{grav} = GM/R^2 = 10^{14}$ cm s$\!^{-2},$
   which (along with the plasma temperature of the order of several keV)
     determines the pressure scale height $\sim 0.1-1$ m,
       exceeds the familiar terrestrial one by eleven orders of magnitude.
 The flow velocity is close to 120000 km s$\!^{-1},$ i.e., it exceeds the speed of sound
   and the velocities of any motions in the Earth's atmosphere by five orders of magnitude.
 It exceeds even the parabolic velocity on the Earth's surface and, hence, the velocity of the rockets
   that place interplanetary vehicles in orbit by four orders of magnitude.
 Even the fastest asteroids and comets had velocities lower by thousands of times
   during their entry into the Earth's atmosphere.

 \subsection{The Levitating Layer}

 In our previous paper (Inogamov and Sunyaev 1999; below referred to as IS99 for short),
   we suggested solving this problem in the 1D approximation by reducing it to the problem
     of the spreading of a thin layer of rapidly rotating matter over the NS surface (Fig. 1).
 We assumed that the main friction took place in an even thinner base of the spreading layer
   and the picture is similar to the braking of a supersonic flow near a wall.
 This picture had its fairly unexpected consequences.
 The energy release during the braking of such a flow is so great that the local radiation flux
   per unit surface area is close to the Eddington value, $q_{Edd} = L_{Edd}/4 \pi R^2 = $
     $m_p g_{grav} c/\sigma_T \sim 10^{25}$ erg s$\!^{-1}$ cm$\!^{-2},$
       while the radiation pressure force in the flow zone easily compensates for the difference
         between the force of gravity and the centrifugal force.
 As a result, the effective acceleration due to gravity determined by the combined action
   of these three forces works.
 Accordingly, the pressure scale height $\sim 1$ km increases sharply compared to $\sim 0.1-1$ m
   (in the absence of a powerful radiation flux and rapid rotation)
     and the latitudinal flow levitates above the stellar surface.

\begin{figure}[t]
 \includegraphics[width=1\columnwidth]{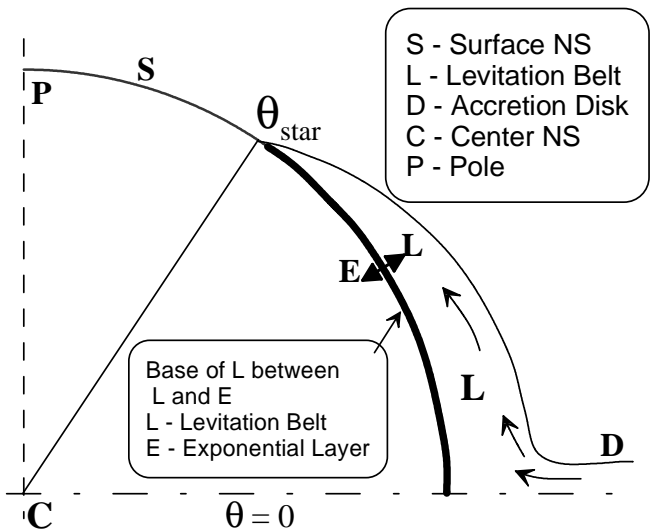}
 \caption{\label{fig:1}
 Accretion of matter from the disk D into the spreading layer L.
 In the spreading layer or belt, the matter levitates (the letter L on the arrow LE).
 The base is indicated by the thick arc extending from the equator $\theta = 0$ to latitude $\theta_\star.$
 This is the zone of contact between the layer L and the neutron star.
 It bounds the belt L from below.
 In the base of the belt L and under it, the density $\rho$ grows quasi-exponentially.
 This is indicated by the letter E (Exponential) on the arrow LE.
 This growth continues up to the boundary between the exponential atmosphere and the ocean
   that forms a layer of degenerate fluid above the solid NS crust surface.
 The NS surface is denoted by the letter S; C and P are the NS center and pole.
      }
\end{figure}

 The meridional component of the centrifugal force is directed toward the equator.
 It prevents the latitudinal flow from expanding along the stellar meridian.
 The flow expansion along the meridian is possible only through the loss of angular momentum
   due to friction on the wall and the corresponding decrease in the meridional component
     of the centrifugal force (see Section 2.8).
 The height of the levitating layer decreases as one recedes from the equator.
 Thus, the meridional pressure gradient maintains the poleward spreading of the matter
   (losing its angular momentum) along the meridian (Fig. 1).

 The upper boundary of the spreading layer along the meridian $\theta_\star$ shown in Fig. 1
   is determined by the area on which the energy release due to flow braking
     can provide a nearly Eddington radiation flux per unit surface area
       and levitation with a high speed of sound and a fairly large pressure scale height.
 The IS99 solution predicts the presence of two equidistant (from the equator) latitude bands
   with the Eddington surface brightness in X-rays on the surface of an accreting NS.
 The distance from the bright bands to the equator and their width increase with accretion rate.
 The levitating belt covers the entire stellar surface at an Eddington accretion rate
   $\dot M \approx 2\times 10^{18}$ g s$\!^{-1}.$
 The low surface brightness near the equator is related to an infinitesimal effective gravity
   due to the closeness of the flow velocity to its Keplerian value.
 The advection of radiation energy (hydrodynamic heat transport) contributes to the removal of the energy
   (being released in the equatorial zone) to higher latitudes.
 The scheme for advective photon transport is shown in Fig. 2.
 The advection of radiative energy is responsible for the formation of the mentioned bright bands.

\begin{figure}[t]
 \includegraphics[width=1\columnwidth]{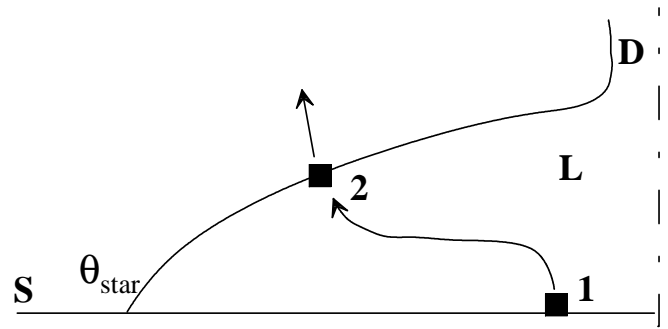}
 \caption{\label{fig:2}
 Hydrodynamic advection of photons along the meridian poleward.
 Radiation is transported by the spreading layer from point 1 to point 2 along the track 1$\to$2
   with energy radiated away from the surface of the levitating belt L at point 2 (see IS99).
 In contrast to the previous Figure, the stellar boundary S is indicated here by the straight line
   and is placed horizontally.
      }
\end{figure}

 The IS99 solutions were based on the conclusion by Popham and Sunyaev (2001) about the presence of a thin neck
   in the accretion disk D at the place of mass transfer from the disk into the spreading layer L (see Fig. 1).
 This disk thinning at the inner boundary (the disk–boundary layer boundary) is a natural consequence
   of the angular momentum conservation law (for a discussion, see Shakura and Sunyaev 1973).

 \subsection{The Base of the Spreading Layer}

 In this paper, we made an attempt to calculate the structure of the "sole" or the base of the levitating layer
   in which (a) the flow brakes, (b) the main energy release takes place,
     and (c) the accreting matter settles
       (the base is the arc $0 <\theta<\theta_\star$ of circumference S in Fig. 1).
 We hoped to get the picture of a thin sublayer near the NS surface that would confirm the assumptions
   made when deriving the IS99 equations.
 Below in this section, we briefly summarize the results of our attempts to find a solution of the problem
   that leads to an efficient braking of the flow under consideration.
 The solution was sought in the context of a well-known approach to turbulent viscosity
   well studied in nature and numerous experiments.
 We are talking about the Prandtl-Karman model of turbulent friction (Schlichting 1965).
 In addition, for checking we used the second approach with $\alpha$-friction
   suggested by Shakura and Sunyaev (1973) for accretion disks.

 Here, we will consider in detail the physical processes in an axisymmetric flow
   that take place in a narrow (along the meridian) column of matter
     (a "pipe"; see Section 1.10 and Fig. 19 below) under the levitating layer.
 This approximation is valid, because the column height is small
   compared to the extent of the base along the meridian.
 Attention will be focused on the decrease in the azimuthal velocity of the matter in this column
   and on the change in radiation flux and matter temperature with depth or, which is more convenient,
     with surface density $\Sigma$ of matter between the point of interest to us at a given depth
       and the lower boundary LA of the levitating layer (see Fig. 3).
 Obviously, the concept of a solid wall is absent in our formulation of the problem --
   the flow of rarefied matter in the levitating layer L rushes above the atmosphere
     made up of the plasma that left this layer slightly earlier.

\begin{figure}[t]
 \includegraphics[width=1\columnwidth]{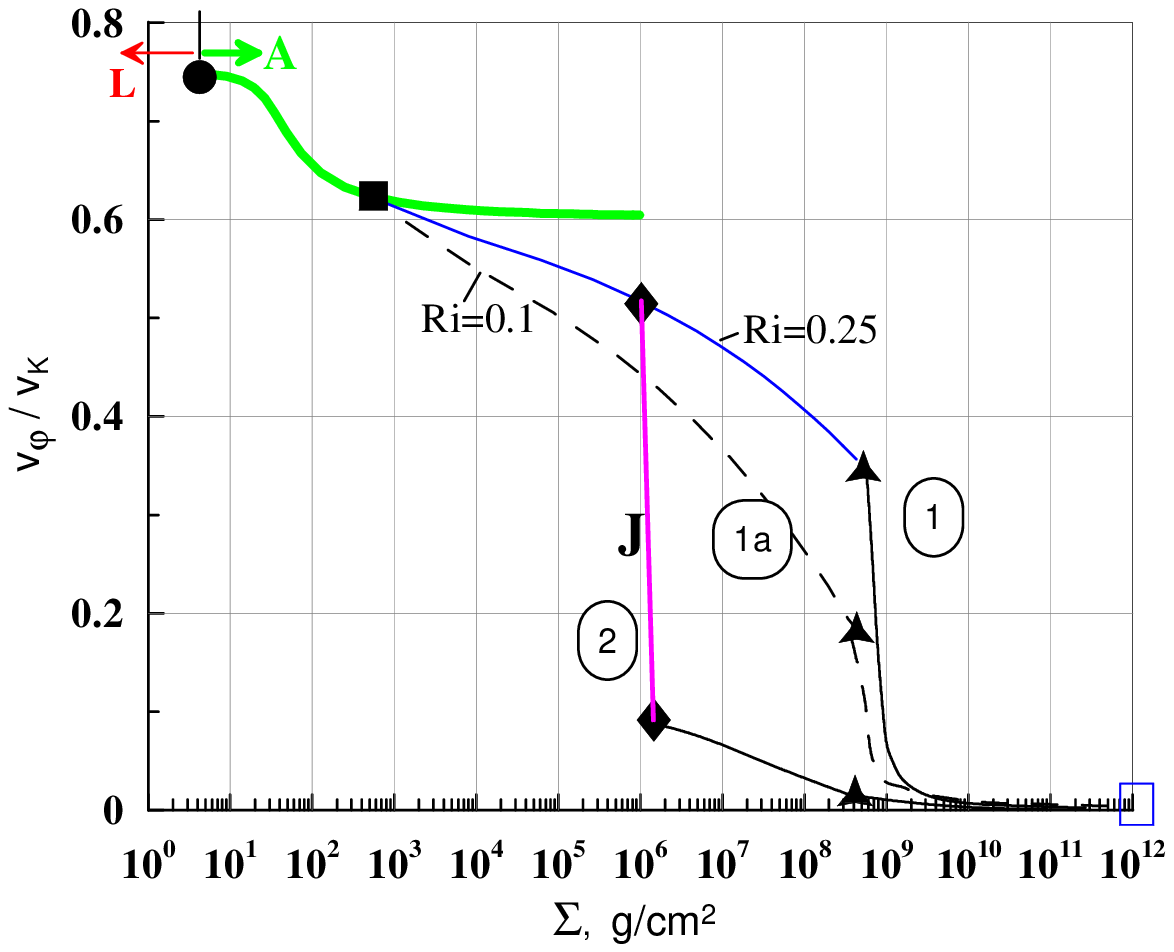}
 \caption{\label{fig:3}
 Rotation velocity profile as a function of column density $\Sigma.$
 Here, $\Sigma$ is measured from the horizon LA
   where matter is balancing on the verge of a wind;
    L denotes the levitating layer; A is the upper part of the base;
     and LA is the boundary between the layers L and A.
 It is highlighted by the filled circle.
 The layer L is to the left of this circle, see arrows in the figure.
 The profiles with labels Ri =0.25 and 0.1 refer to the standard (0.25) and reduced (0.1) values
   of the Richardson number, which is equal to the square of the ratio of the gravitational
     and shear frequencies.
 One can see that even in the case with reduced Ri-number (Ri=0.1) the gravity stratification
   strongly decreases the exchange of momentum due to shear turbulence.
 This gives maximal decrease of $v_\varphi$ with depth.
 The decrease in velocity $v_\varphi$ between the filled circle and square
   is related to turbulent shear friction according to the Prandtl-Karman model (see Section 1.3).
 The gravity wave ladder is located between the filled square and triangle.
 In case 2, the existence of a giant gravity wave is assumed (see Section 1).
 It reduces sharply the velocity $v_\varphi$ from one diamond to another (velocity jump J).
 In cases 1 and 1a, the velocity $v_\varphi$ gradually decreases in the exponential atmosphere
   between the filled square and triangle.
 Unless the rapid rotation is terminated by the jump J, very dense and massive layers fall into rotation
   with high frequencies.
 The triangle corresponds to the helium flash ignition horizon;
   the ocean of degenerate fluid begins under it.
 The ocean occupies the layer between the triangle and the open square
   (this is the upper boundary of the solid crust).
      }
\end{figure}

 {\bf The settling of matter into the base from the levitating layer.}
 In IS99, we neglected the settling of matter from the levitating layer L,
   because this layer spread above an impermeable wall.
 According to this simplest picture, all of the accreting matter from the disk
   was transported along a spiral (the rotation plus the motion along the meridian) to the shock wave
     at the boundary $\theta_\star$ of the layer L (see Figs. 1 and 2).
 Beyond the latitude $\theta_\star$ $(\theta >\theta_\star),$ levitation ceases
   and the cold matter passed into corotation with the star spreads poleward.
 According to IS99, this corotating matter fills the entire NS surface
   and, in particular, forms the underlying envelopes beneath the belt L.

 In this paper (for the equations and a detailed discussion of the solution, see Sections 3--8),
   we take into account the settling of matter inside the levitating belt L.
 The settling rate is specified by the function $\dot \Sigma(\theta).$
 It characterizes the settling in the vertical column of the latitude ring
   at various angular distances $\theta$ from the equator.
 Allowance for $\dot \Sigma(\theta)$ gives rise to the term in the angular momentum equation
   responsible for the radial advection of angular momentum $\propto\dot \Sigma(\theta) v_\varphi$
     (see Section 4.1).
 Similarly, such advection enters into the energy equation.

 The settling rate $\dot\Sigma$ is low near the equator
   (where the rotation velocity of the matter in the levitating flow is nearly Keplerian)
     and is comparable to the advection rate upward along the meridian near the surface brightness maxima
       (above, we have already mentioned the existence of two bright latitude bands predicted in IS99).
 To calculate $\dot\Sigma(\theta)$ at fixed latitude $\theta,$
   in Section 6.5 we use the condition for the transition to a state of levitation at the boundary
     between the belt L and the base.
 The force stresses related to turbulent friction and nonlinear interaction in the gravity wave ladder
   are responsible for the braking of Keplerian rotation.

 {\bf Difficulties with the standard approach to the problem suggesting the flow braking in a thin base
   through turbulent viscosity.}
 A detailed analysis (see Sections 1.3 and 8 in the full paper) showed that the simple solution with a thin base
   and a gradual transition (see Fig. 1, the arrow L$\to$E) from rapid rotation in the levitating belt L
     to slow NS rotation is possible,
       but it imposes stringent requirements on the viscosity parameters in this base.
 These requirements are in conflict with the available experimental data on viscosity
   in near-wall supersonic flows.

 The solution with a smooth rotation velocity profile $v_\varphi(r)$
   in which the velocity $v_\varphi$ decreases to the NS rotation velocity through turbulent viscosity
     exists only when the viscosity coefficient in the Prandtl-Karman model is chosen to be
       hundreds of times less than its universally accepted value (see Section 8).
 Likewise, the $\alpha$-coefficient in the Shakura-Sunyaev (1973) model of viscous stress
   should be $10^{-3}$ at the speed of sound including the radiation pressure
     or $10^{-2}$ if the thermal ion velocity is substituted into the formula for $\alpha$ (see Section 11.2).

 Velocity distributions for different viscosities are shown in Figs. 4-6.
 They describe the decrease of $v_\varphi$ in the upper part of zone A shown in Fig. 3 by the thick curve
   which starts from the point LA and goes down in depth.
 The thick curve in Fig. 3 corresponds to the part of the zone A where the velocity shear dominates the flow.
 Here oscillations due to gravity waves are suppressed
   since the Richardson number
   $$
   {\rm Ri}=N^2/\omega_{sh}^2
   $$
    is small: Ri$\ll 1;$
     $N$ and $\omega_{sh}$- are the Brunt-V\"ais\"al\"a and shear frequencies.

\begin{figure}[t]
 \includegraphics[width=1\columnwidth]{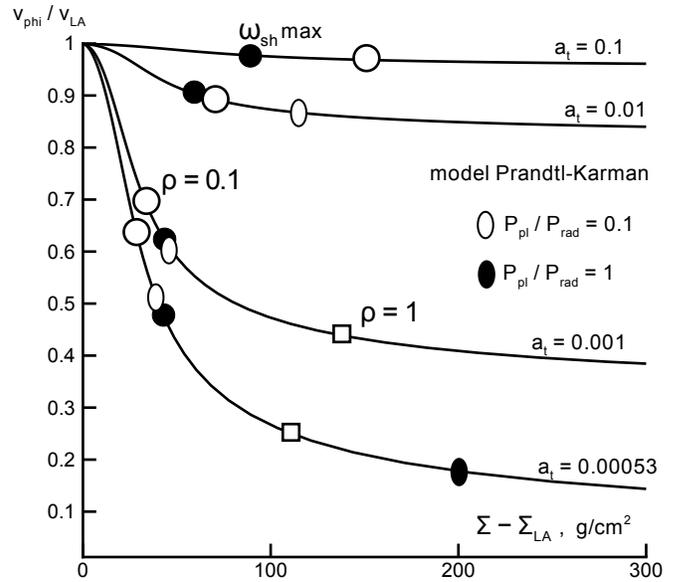}
 \caption{\label{fig:3a}
  Decrease of linear angular velocity $v_\varphi$ and sharp accumulation of column mass $\Sigma$
    in the process of braking of velocity.
  The part of the zone A, where the velocity shear is dominated, is shown.
  This is marked by the thick curve in Fig. 3.
  The filled circles indicate the maxima of the shear frequency $\omega_{sh};$
    the open circles and squares represent respectively the depths where densities are:
      $\rho =0.1$ and 1 g cm$\!^{-3}.$
  The elliptical symbols show the ratio of plasma and radiation pressures.
      }
\end{figure}

\begin{figure}[t]
 \includegraphics[width=1\columnwidth]{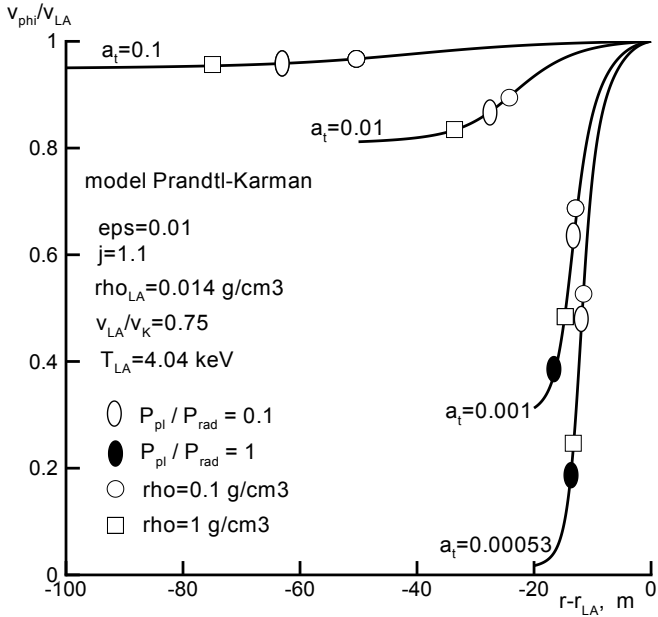}
 \caption{\label{fig:3b}
  Velocity profiles in the Prandtl-Karman model of friction (4.8).
  The influence of the coefficient $\alpha_t$ (the numbers near the curves)
    on the flow braking in the layer A is shown.
  The Prandtl-Karman model is traditionally used in calculations of stress in the turbulent shear flow.
      }
\end{figure}

\begin{figure}[t]
 \includegraphics[width=1\columnwidth]{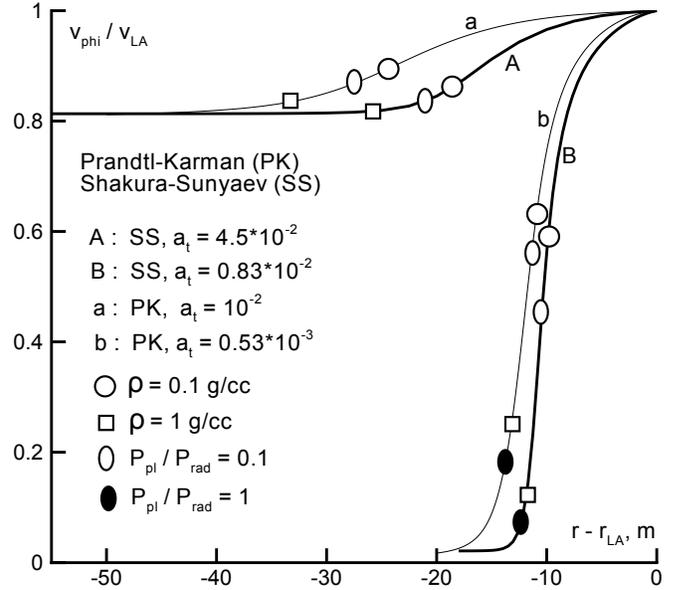}
 \caption{\label{fig:3c}
  Surprising universality of the conclusion about the impossibility
    of stopping a Keplerian flow by shear friction
      at a physically justified friction coefficients
        independent of the specific formula for the turbulent shear viscosity $\nu_t.$
 Profiles a, b and A, B refer respectively to the calculations
   based on the qualitatively different equations (4.8) and (4.9) for viscosity $\nu_t.$
 Equations (4.8) and (4.9) correspond to the Prandtl-Karman (PK) model
   and to the Shakura-Sunyaev (1973) $\alpha$-parameter model.
      }
\end{figure}

 A situation with the critical Richardson number Ri$\approx 0.25$ is shown in Fig. 7.
 Discussion of the critical value may be found in Kippenhahn and Thomas (1978);
   Rosner et al. (2001);
   Durci et al. (2002);
   see also wikipedia http://en.wikipedia.org/wiki/Richardson-number.
 The expression $N$ for plasma with a significant contribution of radiation pressure
   is described in the full paper; $\omega_{sh}=dv_\varphi/dr.$
 In Fig. 3 the Prandtl-Karman model and the most plausible minimal value of $\alpha_t$ is used:
   $\alpha_t=0.01.$

\begin{figure}[t]
 \includegraphics[width=1\columnwidth]{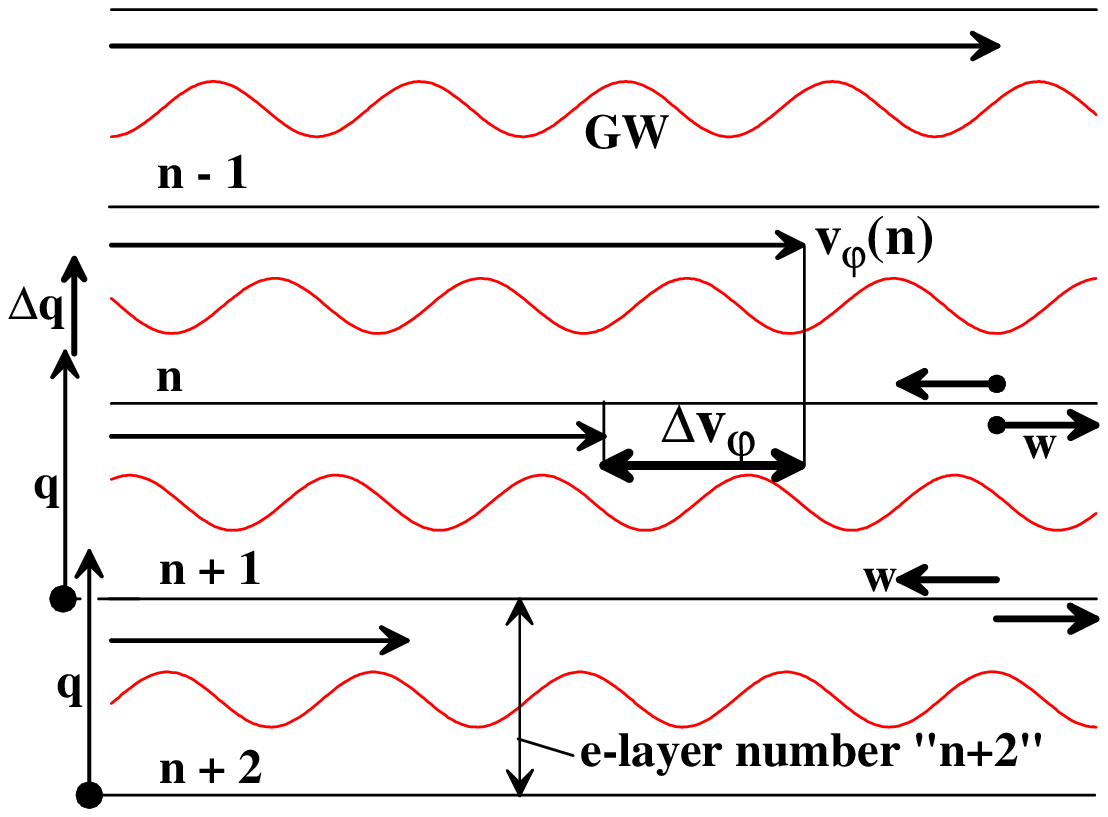}
 \caption{\label{fig:4-7}
  Sequence or ladder of nonlinear gravity waves,
    each traveling in its layer with a thickness of $h_E=2T/m_p g$ of an exponential atmosphere.
  The pairs of arrows $w$ indicate the transferred shear stress
    through which the angular momentum is eventually transported to the bulk of the NS.
  The arrows $v_\varphi(n)$ indicate the average rotation velocity at the level of layer n.
  The arrows $q$ and $q + \Delta q$ represent the radiation flux at the entrance to or exit from the e-layer
    (an increase in $\rho$ by a factor of ``e" takes place in the ``e-layer").
  The increment $\Delta q = w \Delta v_\varphi$ is related to wave breaking
    and the dissipation of kinetic energy of the gravity wave in this layer.
  The temperature $T$ drops by $\Delta T$ when shifting upward by one layer to provide the removal
    of flux $q$ through radiative heat conduction.
      }
\end{figure}

 In the Prandtl-Karman model turbulent viscosity $\nu_t$ is calculated as
 $$
 \nu_t=\alpha_t l_t v_t =
 $$
 $$
 = \alpha_t h_\rho (h_\rho\, d v_\varphi/dr) =
 $$
 $$
 = \alpha_t h_\rho^2 \,d v_\varphi/dr,
 $$
 see equation (4.8) in the full paper.
 Here $\alpha_t$ is a non-dimensional coefficient,
   $l_t$ and $v_t$- are respectively  characteristic spatial scale (average size of vortices)
     and an amplitude of velocity fluctuations.
 Value $h_\rho(r)=[\,|d\ln(\rho)/dr|\,]^{-1}$- is a spatial scale of density distribution $\rho(r).$
 Kinematic viscosity $\nu_t$ allows the calculation of dynamic viscosity $\mu=\rho \nu$
   and viscous stress $w$ via the Newton friction law
   $$
 w = \mu \frac{dv_\varphi}{dr}.
   $$
 The stress between the rotating fast layers and the deeper interiors of NS
   decreases rotation frequency with depth.

 The Prandtl-Karman model is traditionally used in analyzing shear turbulence.
 The coefficient $\alpha_t$ in this model is related to the Karman constant $\kappa$ (Schlichting 1965)
   by the formula $\alpha_t = \kappa^2.$
 In the case of turbulence near a wall in a medium with neutral stratification,
   we have
   $$
   \kappa\approx 0.4, \;\;\; \alpha_t\approx 0.16.
   $$

 In the Shakura and Sunyaev (1973) model of $\alpha$-friction the turbulent viscosity is
 $$
 \nu_t=\alpha_t l_t v_t=\alpha_t h_\rho c_{T}.
 $$
 Again $l_t$ and $v_t$ are the shear turbulence length and velocity scales,
   and $\alpha_t$ is a numerical coefficient.
 The vertical size of the vortices is determined by the stratification scale.
 We can use both the scale $h_\rho$ (density stratification)
   and the scale
   $$
   h_{ppl} =[\,|d \ln (p_{pl)}/dr|\,]^{-1}
   $$
    (thermal plasma pressure $p_{pl}$ stratification).
 Since the change in temperature is small when varying the radius,
   these definitions are approximately equivalent.
 In cases, when radiatin pressure dominates,
   the scale $h_p$ calculated from the total pressure can be considerably larger.

 Our objective is to estimate the maximum possible decrease in rotation velocity $(\Delta v_\varphi)_A$
   on the layer A.
 This is the layer in Fig. 3 between the horizon LA and the horizon of the jump J.
 Such an estimation requires the minimal, physically justified viscosity $\nu_t.$
 Therefore, first, we use the length $h_\rho$ rather than $h_p$
   and, second, we use the smallest pulsation velocities $v_t.$
 Let us discuss the choice of the characteristic velocity $v_t$ in the Shakura-Sunyaev model.
 The flow rotational velocity is great, $v_\varphi \sim v_K$
   $\approx 0.4 \, c$ $\approx 10^{10}$ cm s$\!^{-1}.$
 The adiabatic speed of sound $c_S =(0.1-0.2) v_\varphi \sim 10^9$ cm s$\!^{-1}$
   in the radiation-dominated medium under study is also fairly high (IS99).
 The subscripts S and T refer to the adiabatic and isothermal speeds of sound.

 In ordinary turbulent flows at great Reynolds numbers, the velocity of turbulent pulsations
   is several percent of the mean flow velocity (e.g., 5-10\%),
     then $v_t \sim 0.07 v_\varphi \sim 7 \times 10^8$ cm s$\!^{-1}.$
 The isothermal plasma speed of sound $c_T \sim 10^8$ cm s$\!^{-1}$ under our conditions
   is $\sim 1\%$ of the flow velocity.
 Ion thermal velocities are of the order of $c_T.$
 As one can see, this is the lowest velocity.
 Thus, the smallest scales $l_t$ and $v_t$ are used in the approximation of viscosity $\nu_t.$

 From what was said about the turbulent pulsation velocities $\sim 7\times 10^8$ cm s$\!^{-1}$
   it follows that the coefficient $\alpha_t$ in the Shakura-Sunyaev viscosity should be $\sim 1.$
 But as we see from Fig. 6 much smaller coefficients are necessary to stop fast rotation
   in the velocity shear layer.

 Under standard assumptions about the parameters of turbulent viscosity,
   it can reduce only slightly the velocity of the very rapid nearly Keplerian rotation of the flow
     that slowly settles onto the NS (see Section 8 in the full paper).
 We arrive at a paradoxical result:
 {\it the universally accepted parameters of turbulent viscosity
   give rise to a massive rotating equatorial layer in which the rotation decays very slowly
     and the energy release takes place mainly at depths
       where the matter density exceeds considerably $10^4$ g cm$\!^{-3}.$ }

 Slow decay of the rotation velocity causes the radial component of the centrifugal force to decrease.
 For the same reason, the radiation flux
   produced through energy release due to the decay of the rotation velocity
     in the layers below that under consideration also decreases with depth.
 Both these factors cause a rapid increase in "effective" gravity with depth
   and a sharp decrease in pressure scale height.
 It is important to note that a slight imbalance between the three forces under consideration
   (gravity, centrifugal force, and light pressure)
     is enough for the local pressure scale height $h_E=2T/m_p g_{eff}$
       to begin to decrease rapidly with depth
         and for the matter density and its contribution to the total plasma and radiation pressure
           to begin to grow sharply (for a detailed analysis, see Sections 1.3 and 8).

 {\bf The gravity wave ladder.}
 The ``sole" or base extends to many heights such as $h_E,$
   i.e., the density in it increases by many orders of magnitude.
 The difference between the velocities of the azimuthal flow above and below a layer
   with a thickness of the order of $h_E$ should lead to the generation of internal gravity waves
     propagating inside the layer under consideration in the direction of the flow.
 The wind above the ocean surface on Earth generates the universally known waves in exactly the same way.
 Nonlinear processes (wave braking) dynamically couple the waves and shear flows (differential rotation)
   in neighboring layers with a thickness of $h_E.$
 This makes it possible to transport angular momentum to deep NS regions through a stably stratified atmosphere
   made up of a ``stack" or ``ladder" of layers $h_E$ (see Fig. 7).

 We constructed the gravity wave ladder by assuming that a small, compared to the Keplerian velocity,
   but noticeable decrease in rotation velocity
     (a decrease by a value of the order of the local speed of sound)
       occurs in each layer with a thickness equal to the pressure scale height $h_E$ (see Fig. 7).
 Unfortunately, full flow braking at the commonly assumed Richardson number (see Sections 1.3 and 10) Ri = 0.25
   requires a large number of ladder "steps" (here, a layer of thickness $h_E$ is called the ladder step).
 Increasing the shear frequency $dv_\varphi/dr$ by reducing the critical value of Ri from 0.25 to 0.1
   does not save the situation in the sense of an overly high mass drawn into rapid rotation (see Fig. 3),
     although, of course, increasing the frequency $dv_\varphi/dr$
       allows the rotation velocity to be reduced with depth more rapidly.
 This approach leads to a picture with great energy release at depths
   corresponding to densities $\rho> 10^4$ g cm$\!^{-3}$ and a mass $\Sigma$ in a unit column
     greater than $10^7$ g cm$\!^{-2}$ (see Fig. 3 and a detailed discussion in Section 1).

 At depths with a surface density $\Sigma > 10^9$ g cm$\!^{-2},$ the matter density rises to an extent
   that the Fermi energy of electrons begins to exceed the plasma temperature
     and the electrons in the plasma become degenerate.
 The density below this layer (that is in the "ocean" of degenerate electrons and thermal ions)
   ceases to grow exponentially with depth, the entropy ceases to decrease,
     and conditions for full braking of the azimuthal rotation are created.
 Here, we are dealing with a picture with a thick, rapidly rotating base.
 The presence of a lower ocean-solid crust boundary again makes the problem similar to the problem
   of flow braking near a wall accompanied by the formation of a logarithmic Prandtl–Karman velocity profile
     (which was considered in detail for the problem of a levitating layer in IS99).
 The solid crust rotates with the NS rotation frequency.

\subsection{Main equations}

 We use equation of mass conservation
 $$
 \rho v_r = - \dot\Sigma
 $$
 and radial force balance
 $$
 \frac{d}{dr} ( p + \rho v_r^2 ) = - \rho\left(g_{grav} - \frac{v_\varphi^2}{R}\right)=-\rho g_{eff}
 $$
 to describe braking, cooling, and settling of accreting matter.
 Here
 $$
 \dot\Sigma\sim \Sigma_T\, c/R,
 $$
  is accretion rate per unit of NS surface,
  $\Sigma_T=m_p/\sigma_T,$ $\sigma_T$ is Thomson cross-section, $R$ is radius of neutron star,
 $v_\varphi$ and $v_r$ are rotational and radial components of velocity.
 The last corresponds to settling of matter.

 Equation of state is
 $$
p = p_{pl}+p_{rad},\;\;\;\;\;
p_{pl}= \rho \frac{2T}{m_p},\;\;\;\;
p_{rad}=\frac{a T^4}{3},
$$
 where $p_{pl}$ and $p_{rad}$ are plasma and radiative pressures.

 Conservation of angular momentum gives
 $$
 w = J - \dot\Sigma \; v_\varphi.
 $$
 This conservation law follows from equation
 $$
 \frac{\partial(\rho \,v_\varphi)}{\partial t}=
 - \frac{\partial(\rho\, v_\varphi\,v_r)}{\partial r}+\frac{\partial w}{\partial r}
 $$
 in the steady-state case when $\partial_t=0.$
 The constant
 $$
 J\sim \dot\Sigma\, v_K
 $$
  is an integral of this equation; here $v_K$ is Keplerian velocity.

 Newton friction law for turbulent viscous stress is
 $$
w=\rho \,\nu \, d v_\varphi/dr,
$$
 where $\nu$ is total kinematic viscosity
 $$
 \nu = \nu_i + \nu_{rad} + \nu_t.
 $$

 Partial viscosities are
 $$
 \nu_{rad} = \frac{4}{15} \,\frac{a \,T^4\, \Sigma_T}{c\,\rho^2}
 \simeq \frac{\rho_{rad}}{\rho} \, l_T \, c
$$
 (Weinberg, 1972), here $\rho_{rad}=a T^4/c^2$ is ``density" of photons,
   $l_T=1/n \sigma_T $ is photon mean free path and $n=\rho/m_p.$
 Ion viscosity is
$$
\nu_i = 10^3 \,\frac{ T^{2.5}}{\rho \, \ln \Lambda},
$$
 where $\Lambda$ is Coulomb logarithm (Spitzer, 1962),
 temperature is in keV, while density and kinematic viscosity are in g/cc and cm$\!^2/$s.
 Under the physical conditions in the spreading layer the radiative viscosity often exceeds the ionic one,
 but the turbulent viscosity dominates.

 Turbulent viscosities $\nu_t$ in case of shear flow with weak stratification
 has been given above.

 In the case of insignificant gravity stratification
 the braking of rotation as result of viscous deceleration is
 $$
 \frac{d v_\varphi}{dr}=\frac{1}{\rho}\,\frac{w}{\nu}=
  \frac{1}{\rho}\,\frac{J-\dot\Sigma\, v_\varphi}{\nu_i + \nu_{rad} + \nu_t}.
 $$

 In the alternative case of significant gravity stratification
 we use another description of the deceleration of rotation (decrease in shear frequency $v'_\varphi).$
 This another description is
 $$
 \frac{d v_\varphi}{dr} = \frac{N}{\sqrt{{\rm Ri}}}.
 $$
 Here $N$ is the Brunt-V\"ais\"al\"a frequency and Ri is Richardson number.
 This frequency in the case of large contribution of radiation pressure $p_{rad}$
 in the layer transmitting radiative energy flux $q$ is
$$
N^2 = g_{eff}^2 (c_T^{-2}-c_S^{-2})
\left( 1 - \frac{1}{2} \;\tilde q \;
\frac{10 + 8 \tilde R + 5/(4 \tilde R)}{1 + 4 \tilde R} \right),
$$
 (see Appendix in the full paper).
 Here ratio $\tilde R$ and normalized flux $\tilde q$ are
 $$
  \tilde R=\frac{p_{rad}}{p_{pl}},
  \;\;\tilde q=\frac{q}{g_{eff} c \Sigma_T}.
  $$

 The deceleration $v'_\varphi$ changes from $v'_\varphi=w/\rho\nu$ (if $0<{\rm Ri}<0.25)$
 to $v'_\varphi=N/\sqrt{{\rm Ri}}$ (if ${\rm Ri}\approx 0.25)$ depending on local value of the number Ri.

 Energy balance
 $$
-\dot\Sigma \,\left(
\frac{ v_r^2 + v_\varphi^2}{ 2} +
5 \frac{T}{m_p} + \frac{4}{3} \, \frac{a T^4}{\rho} +
\psi \right)
- w v_\varphi + q = Q
$$
 presents advection of rotational and radiation energies
$\dot\Sigma (v_\varphi^2/2 + 4\, a\, T^4/3 \rho), $
 viscous dissipation $w v_\varphi,$ and radiative flux
 $$
q = - \frac{c \, \Sigma_T}{3\rho}\frac{d(a \, T^4)}{dr}.
 $$
 They are the main contributions to the energy balance.
 The terms connected with advection of kinetic energy of radial motion $v_r^2/2,$
 thermal energy of plasma $T/m_p,$
 and gravitational energy $\psi$ are small (nevertheless all they are included into solution).
 Changes in gravitational energy appears due to motion of matter in radial direction.
 Radial shifts are small therefore the term $\psi$ is negligible.
 Value $Q$ is the constant (integral) of integration.
 It is defined by boundary condition at large depth.
 Value of $Q$ is connected with energy flux coming from the large depth.
 This value is small in comparison with the main terms listed above.

 \subsection{The Existence of X-ray Bursters Argues against a Massive Base}

 Weak braking leads to slow decay of the azimuthal velocity as one recedes from the levitating layer downward
   along the radius.
 Therefore, significant heat release at great depths is associated with it.
 Note that the store of energy in a nearly Keplerian azimuthal motion exceeds the energy
   being released not only in pycnonuclear reactions but also even during the burning of accreting hydrogen
     to produce helium by more than an order of magnitude.
 Figure 8  
   shows that slow braking (without any abrupt decrease in velocity $v_\varphi)$
     leads to a high temperature (up to 150 keV) at great depths.
 The temperature rises inward to provide the outward removal of the energy being released,
   i.e., to produce a temperature gradient capable of maintaining the corresponding radiative energy flux
     (see Figs. 3,
     8,  
      and 9). 
 This energy release is much greater than the heating through contraction as the accreting layer
   corotating with the star
     settles down.

\begin{figure}[t]
 \includegraphics[width=1\columnwidth]{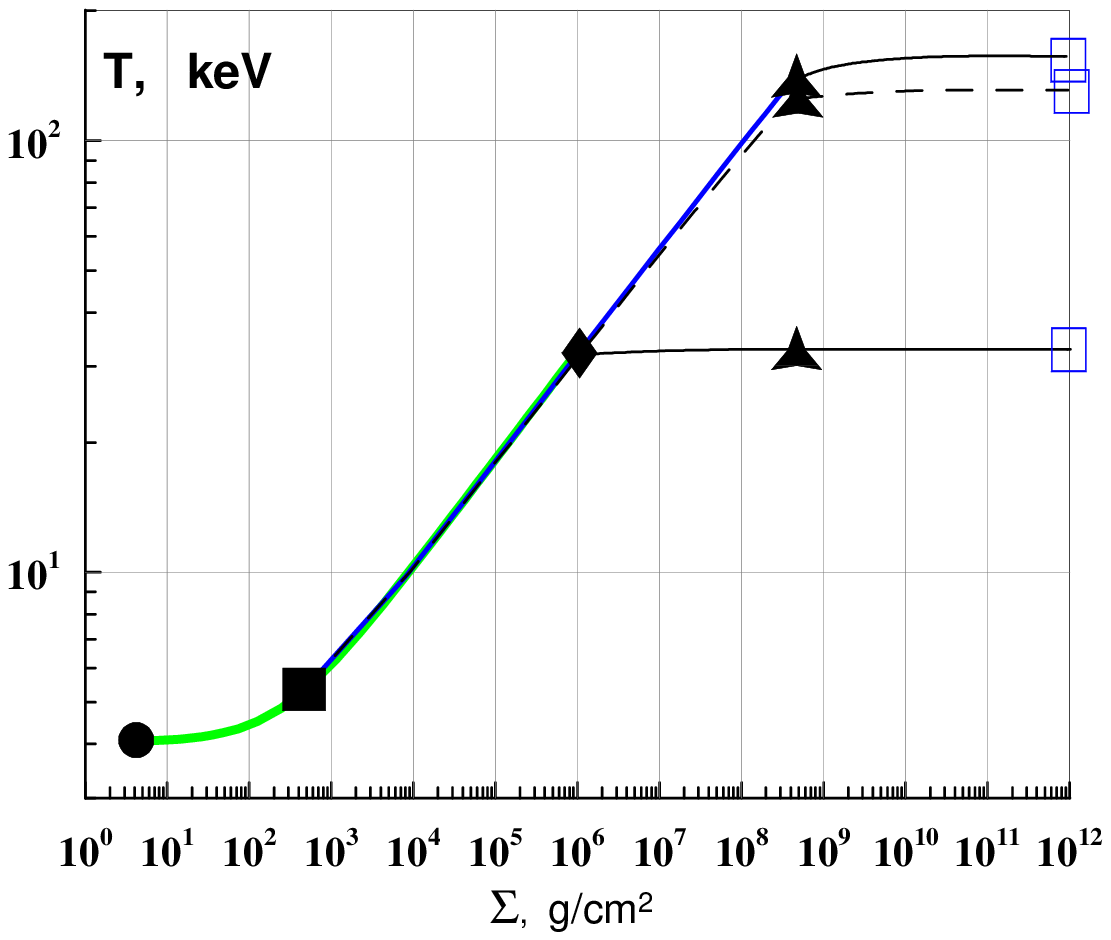}
 \caption{\label{fig:5-8}
  Rise in temperature $T$ deep into the base in our calculations
    with the Richardson number Ri =0.25 (upper curve), with Ri =0.1 (middle curve),
      and for the case with a velocity jump J (see Fig. 3; in Fig. 3 the jump J lies between the diamonds)
        (lower curve).
 In the first two cases, the rotation decays at large depths (see Fig. 3) with a great optical depth.
 This causes a dramatic rise in temperature on the horizon of nuclear helium flashes marked by the triangle.
 The transition from the exponential atmosphere to the ocean of degenerate fluid
   is located approximately on the same horizon.
 The diamond corresponds to jump J; the change in friction mechanism from turbulent viscosity
   to a gravity wave ladder is marked by the filled square;
     the open square indicates the location of the upper boundary of the solid NS crust.
      }
\end{figure}

\begin{figure}[t]
 \includegraphics[width=1\columnwidth]{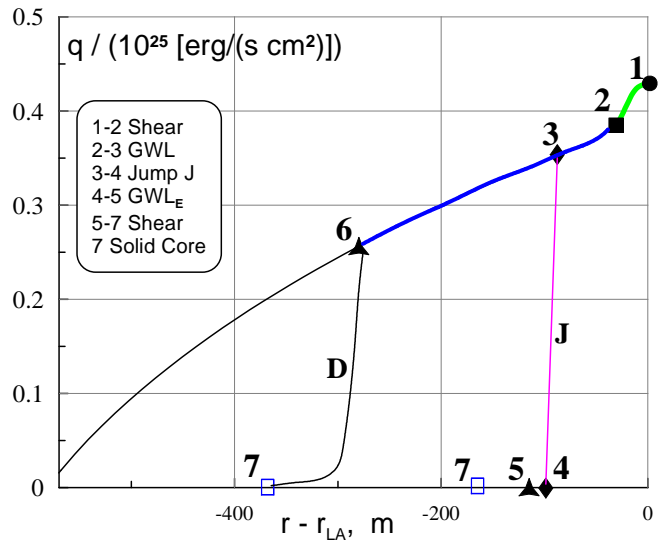}
 \caption{\label{fig:6-9}
  Profiles of the radiation flux $q$ for the cases with a jump in velocity $v_\varphi$
    and with a gravity wave ladder.
 In the presence of jump J, the flux $q$ terminates immediately under the jump
   due to a dramatic slowdown of the rotation velocity $v_\varphi.$
   The flux $q$ is produced by the dissipation of rotational kinetic energy.
   The markers denote the following: circle 1 -- the boundary LA of levitation on the verge of a wind
     (the layer L is to the right of point 1);
       2 -- the change in friction regime;
         3 -- the horizon on which jump J can be located;
           4 -- the horizon under the jump;
             5 -- the atmosphere-ocean boundary;
               7 -- the crust (the crust position is different for different profiles,
                 because it is measured from boundary LA);
                   6 -- the position of the atmosphere-ocean horizon in the case of a gravity wave ladder
                     without any jump J.
 The profile 1-2-3-4-5-7 contains a jump.
 There are the segment of turbulent friction 1-2, the ladder 2-6 with Ri =0.25,
   and the second segment of turbulent friction in the ocean 6-7 on the profile 1-2-3-6-7.
 The profile on which $v_\varphi$ is nulled at a depth of 600 m was constructed
   by assuming the absence of an ocean when the exponential atmosphere extends to the depth of 600 m.
 The distribution $q(r)$ and the optical depth define the distributions of the radiation energy density
   and temperature in Fig. 8.   
      }
\end{figure}

 According to the simplest one-dimensional models (disregarding the presence of a spreading layer),
   at a high rate of spherically symmetric accretion,
     the temperature in a zone with a surface density $\Sigma \sim 10^9$ g cm$\!^{-2}$
       turns out to be so significant that helium in it burns stationarily rather than explosively,
         i.e., without giving rise to X-ray bursts (for a review, see, e.g., Strohmayer and Bildsten 2006).
 Disk accretion and the presence of a spreading layer with a thick base
   should also lead to a dramatic temperature rise in the helium burning zone
     and to the switch-off of X-ray bursts (see Fig. 8). 
 Moreover, the mechanism with a long "ladder" and a great spun-up mass
   leads to a temperature rise on the flash nucleation horizon even at moderate accretion rates,
     when the stationary luminosity of the X-ray source is $\sim 10\%$ of $L_{Edd}.$
 Most of the bursters are active precisely at these persistent luminosities of the source.

 Quasi-periodic oscillations of the radiation flux with a frequency close to the NS rotation frequency
   are observed during intense bursts (Strohmayer and Bildsten 2006).
 This is direct evidence that the base rotating with a high frequency in these sources
   does not reach the helium burning zone.
 In this case, the burning zone rotates with a frequency close to the rotation frequency
   of the central NS regions.

 There is no doubt that the mechanism for the operation of X-ray burst sources is fairly complex --
   most bursters are known to exhibit bursts by no means constantly.
 Series of bursts or single bursts are observed from many of them very rarely
   and occasionally recur only tens of years later
     (see, e.g., Linares et al. (2010) for the source Circinus X-1).
 It is quite possible that in some of these sources we are dealing with a thick base with a high temperature
   in the helium burning zone, because of which no bursts are usually observed in them.
 In a layer with $\rho\sim 10^5 - 10^6$ g cm$\!^{-3},$ a temperature of $\sim 50$ keV
   is enough for stationary helium burning to take place without any nuclear explosions
     that lead to the observed phenomenon of X-ray bursters.
 However, when a burster begins to manifest itself, we must recognize that nature finds ways
   to brake the rotation of the accreting matter in a layer
     located well above the zone of explosive helium burning.

 \subsection{Giant Solitary Gravity Wave As a Way to Stop the Rotation in the Base}

 The analysis performed below in Section 1 leads to a picture with a giant solitary gravity wave (GSGW)
   at a depth that corresponds to a surface density of $10^5 - 10^6$ g cm$\!^{-2}.$
 Such a wave can have the shape of a solitary structure on the equatorial perimeter $[0, 2\pi],$
   which is shown in Fig. 10,  
     or can be a set of several non-equidistant crests on the perimeter.
 In this case, there are crest fluctuations.
 Because of the latter, when averaged over the azimuthal angle $\varphi$ from 0 to $2\pi,$
   the amplitude of the first azimuthal Fourier harmonic is nonzero.
 Such a nonlinear wave brakes the rotation of the matter that falls from the levitating layer L.
 The ratio of the wave amplitude to the pressure scale height $h_E$ and not to the wavelength
   is a measure of nonlinearity.
 The GSGW efficiently interacts both with the rapidly rotating gas in the overlying layer
   and with the underlying layer that rotates considerably more slowly than it.
 Owing to this hydrodynamic interaction, the force accelerating the rotation of the NS as a whole
   is transferred downward to the solid crust. These issues are discussed in detail in Sections 1.3 - 1.6.

\begin{figure}[t]
 \includegraphics[width=1\columnwidth]{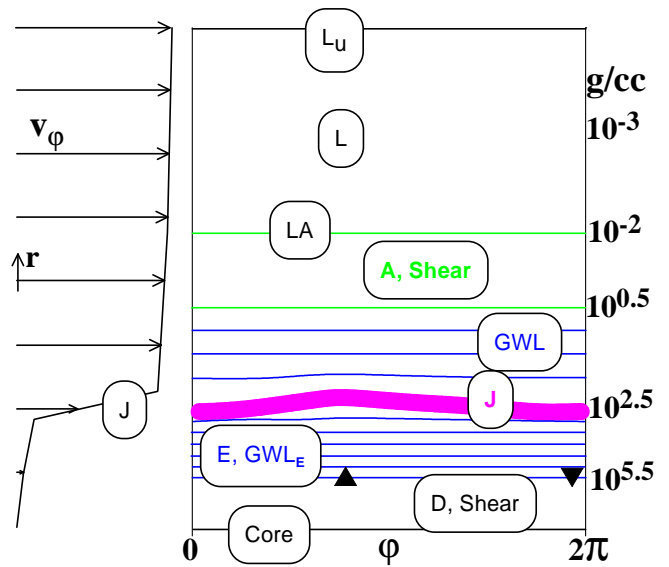}
 \caption{\label{fig:7-10}
 Velocity jump J and giant solitary gravity wave (GSGW) on the period $[0, 2\pi]$
   of the equatorial perimeter $\varphi.$
 The extent of the perimeter is 60 km. The velocity and density [g cm$\!^{-3}]$ profiles
   in the corresponding layers are shown on the left and the right, respectively.
 The designations of the layers and boundaries between them are listed below on the scheme
   of the vertical structure in Fig. 13.  
 The gravity wave ladders are schematically shown in the form of a sequence of e-layers, as in Fig. 7.
 The drop in velocity $v_\varphi$ occurs at the broadened GSGW boundary shown in the form of a wide stripe.
 The main heating of photons takes place in this stripe smoothed by mixing of gases by oblique shock waves.
 The horizontal and vertical scales differ greatly -- the GSGW steepness is less than $10^{-3}.$
 The triangles at level ED (at which a thermonuclear flash ignites) mark the GSGW maximum and minimum
   along the $\varphi$-axis.
 They are important for the explanations that associate the rise in pressure under the GSGW
   with the initiation of a helium flash at a point on the NS surface (see Section 1.8).
      }
\end{figure}

 Such a long-lived feature in the flow can emerge in a rapidly rotating atmosphere,
   just as long waves are generated by a continuous wind during floods in St. Petersburg.
 It is important that this hypothesis can remove the difficulty noted above in the theory of X-ray bursters
   by greatly reducing the stationary temperature in the helium burning layer (see Fig. 8)  
     and opening the possibility of explosive helium burning.
 At present, this hypothesis seems a more reasonable solution than the assumption about anomalously low values
   of the viscosity coefficients
     (hundreds of times lower than those observed in the Earth's atmosphere and ocean
       and in large-scale laboratory experiments)
         or about a sharp decrease (few times) in Richardson coefficient
           when the ladder of interacting gravity waves is considered in a burster's exponential atmosphere.
 With this publication, we hope to draw the attention of specialists in wave dynamics
   to this very interesting and as yet incompletely solved problem.

 {\bf The explosive helium burning ignition zone is located on the equator.}
 In the picture with a GSGW, the temperature under it turns out to be considerably higher
   than that in the commonly considered example of the accretion of matter
     without any angular momentum onto a neutron star (Strohmayer and Bildsten 2006).
 Consider the case of a thin base with a GSGW.
 Let us single out two horizons: the horizon on which the GSGW is located
  $(\rho \sim 10^3$ g cm$\!^{-3},$ $\Sigma \sim 10^6$ g cm$\!^{-2})$
    and the helium flash horizon ED between the exponential atmosphere E and ocean D (see Fig. 11). 
 At the level ED, we have $\rho \sim 10^6$ g cm$\!^{-3}$ and $\Sigma \sim 10^9$ g cm$\!^{-2}.$

\begin{figure}[t]
 \includegraphics[width=1\columnwidth]{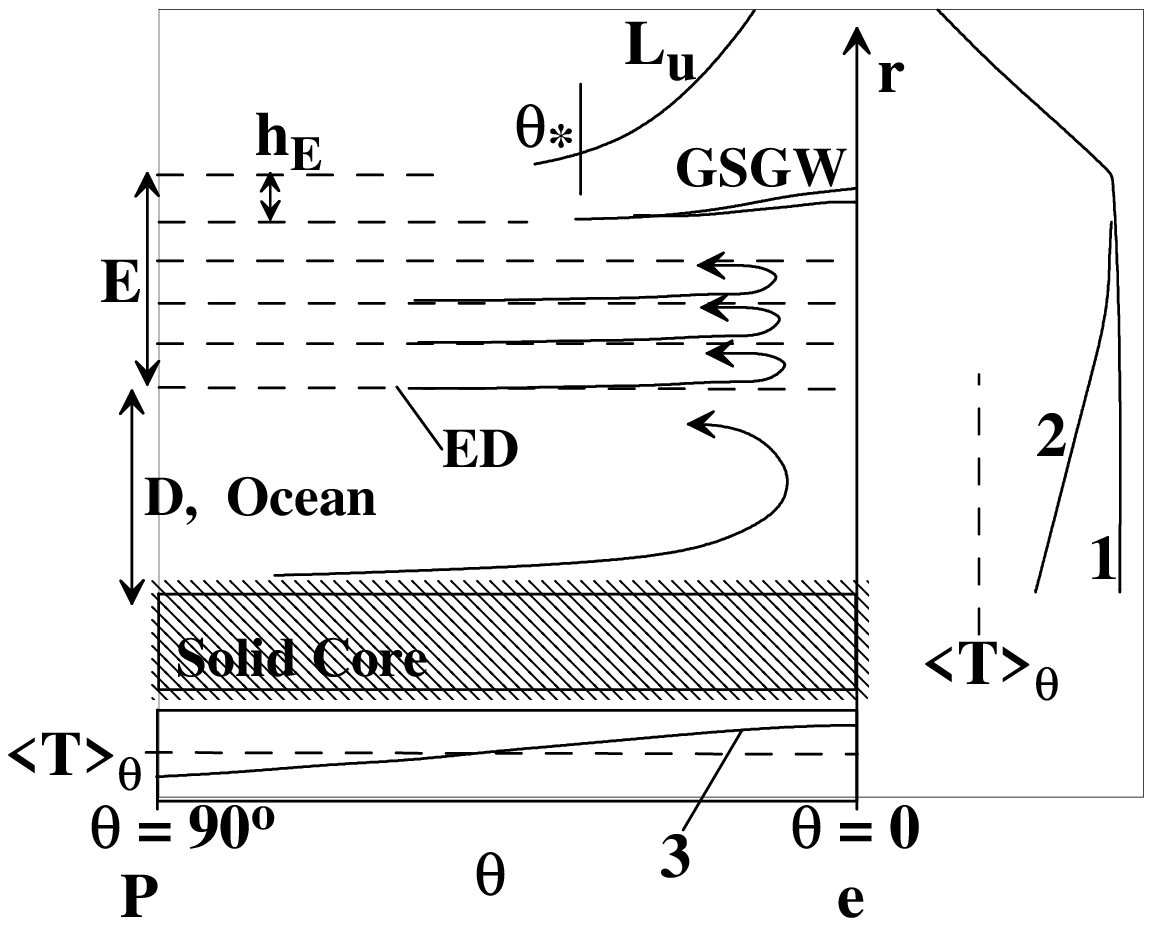}
 \caption{\label{fig:8-11}
 Thermal matching of the zone of GSGW energy release and high-density NS envelopes.
 The meridional NS section and the radial temperature profile $T(r)$ on the equator
   (curves 1 and 2 on the right) as well as the meridional temperature profile $T_{ED}(\theta)$ (curve 3 below)
     are shown.
 Curve 1 was constructed without allowance for the lateral losses.
 These losses were taken into account in curve 2.
 The lateral energy losses from the radial column were caused by meridional convective circulation.
 The temperature at the level ED averaged over the meridian $\theta$ is denoted by $\langle T\rangle_\theta.$
 It is marked by the dashed straight lines on the right and bottom plots.
 The high-density envelopes include: atmosphere E, ocean D, and the solid crust.
 In addition, the photosphere $L_u$ (the zone above which the optical depth is close to unity)
   above the levitating belt L and the boundary of the belt L in latitude $\theta_\star$ are shown;
     P and e are the NS pole and equator, respectively.
 The loops indicate convective circulation in ocean D and atmosphere E.
 In the atmosphere, circulation takes place in layers with a thickness
   equal to the pressure scale height $h_E.$
      }
\end{figure}

 The temperature rise under the GSGW is related to rapid dissipation of the azimuthal velocity
   and great energy release near the GSGW, where the velocity jump is located (see Figs. 3 and 9). 
 The temperature constancy downward along the radius (in the zone where there are no energy sources)
   is the result of energy conservation during the transport of radiation in the radial column
     without any lateral losses.
 Such a temperature profile is established in a time much longer than the diffusion time of the radiation
   from the layer under consideration.

 The GSGW is located at equatorial latitudes,
  where the velocity $v_\varphi$
   (in the levitating belt and the base under it) is at a maximum.
 Strong energy release in the equatorial zone in the simplest approximation of a radial column
   without any lateral losses should lead to a high temperature in the explosive helium burning zone.
 A high temperature makes this zone favorable for the ignition of nuclear reactions of rapid helium burning.

 Let us try to take into account the lateral losses in the radial column of matter and their influence
   on the dependence of the temperature $T_{ED}(\theta)$ on the helium burning horizon ED
     on meridional angle $\theta.$
 The result of our estimations is schematically shown in the lower inset in Fig. 11.  
 Below the horizon of jump J and the GSGW, the matter is in corotation with the NS
   (envelopes E, D, and the crust in Fig. 11).  
 Because of the very high density and, accordingly, heat capacity of the matter in the exponential atmosphere
   and in the ocean with degenerate electrons, the lateral losses are significant
     (the speed of sound depends weakly on radius, while the density rises by several orders of magnitude).
 The ratio of the densities on the horizons of jump J and helium flash ED
   is approximately three orders of magnitude.
 At such a density ratio, lateral cooling causes a decrease in temperature at level ED under jump J
   compared to the temperature immediately under the jump $(\approx 30$ keV, see Fig. 8).  
 This decrease is seen on curve 2 on the right in Fig. 11  
   (cf. curves 1 and 2).
 Level ED is highlighted by the triangle in Fig. 8,  
   while the level of jump J is marked by the diamond in Fig. 8.  

 Despite the presence of lateral losses, the function $T_{ED}(\theta)$ (curve 3 in the lower inset in Fig. 11)  
   has a maximum on the equator (because the heating source is located above it).
 For this reason and since the optical depth $\tau (\theta)\propto \Sigma(\theta)$ above horizon ED
   is at a maximum on the equator (see Fig. 12), the flash will begin on the equator (see Section 1.7).
 The value of $\tau$ above the flash horizon is very important in the theory of a thermal explosion
   when the competition between heating and cooling leads to an explosion.
 The heating is related to the production of heat in reactions,
   while the cooling is related to the radiative heat removal through the optical depth.
 Clearly, poor heat removal contributes to the explosion.

\begin{figure}[t]
 \includegraphics[width=1\columnwidth]{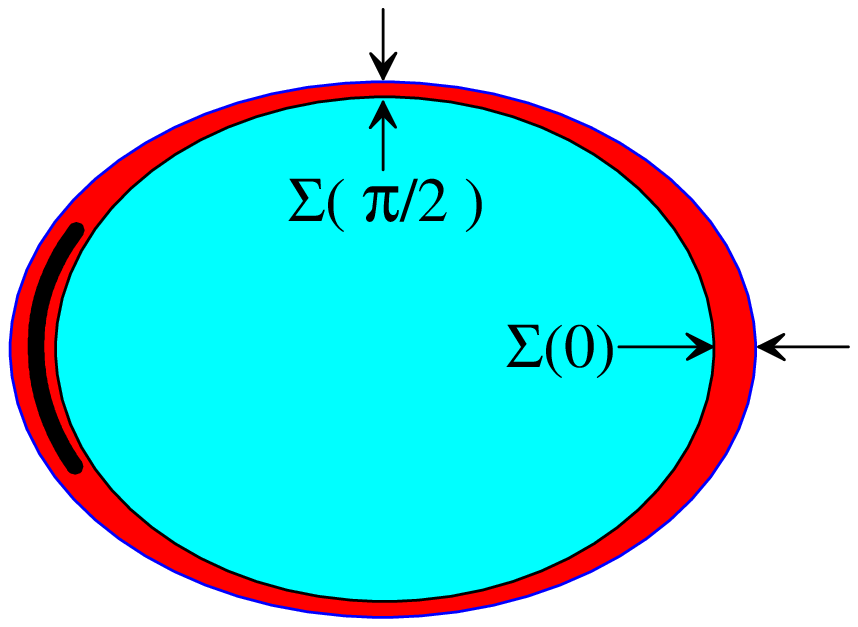}
 \caption{\label{fig:new-12}
 The red layer is the layer of fresh accreted material (fuel) corotating together with neutron star.
 Fresh material is less dense than underlying ashes (blue) from previous bursts.
 The burst begins at the interface between red and blue.
 On rotating star the mass thickness $\Sigma(\theta)$ [g/cm$\!^2]$ of fresh material is function of latitude $\theta.$
 The $\Sigma$ is thicker at equator: e.g., $\Sigma(0)/\Sigma(\pi/2)-1=1.5\cdot 10^{-2}$
   for rotation frequency of star 200 Hz.
 Heat conduction is inversely proportional to $\Sigma.$
 Therefore at equal power of heat production per unit of surface,
  temperature is higher under thicker layer.
  This heat is produced by thermonuclear reactions at the stage preceding the approaching instant of beginning of burst.
  Therefore a thermonuclear burst starts at equator.
  The black curve at the left side of the red layer marks the horizon of giant solitary gravity wave (GSGW).
  Radial thicknesses of fresh sublayers above and below the horizon do not correspond to those
   shown in the sketch.
  Geometrically the above sublayer is thicker, see Fig. 14.
  Above the horizon the fresh material rotates faster.
  But thickness $\Delta\Sigma$ above the horizon is much less $(\sim 10^{-4})$
   than $\Sigma$ of whole fresh layer.
   The addition $\Delta\Sigma$ is insignificant for the ratio $\Sigma(0)/\Sigma(\pi/2).$       }
\end{figure}

 The burst begins in the point at the equator, see Fig. 13.

\begin{figure}[t]
 \includegraphics[width=1\columnwidth]{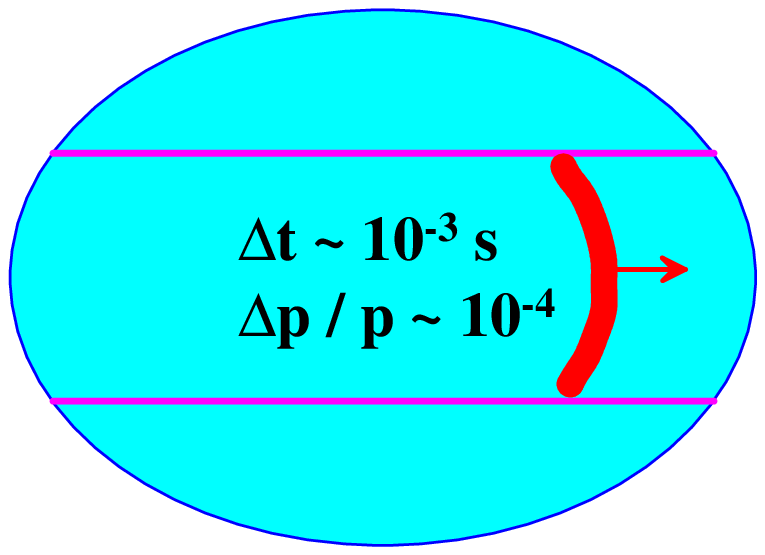}
 \caption{\label{fig:new-13}
 Gathering of thermonuclear fuel (fresh material) at equator (see Fig. 12) and initiation of burst at equator
 under the crest of giant solitary gravity wave (GSGW).
 The crest of GSGW is shown by the red curve.
 It is marked also by the triangular directed up in Fig. 10.
 GSGW propagates along the arrow.
 It propagates above the red/blue surface corresponding to the bottom of the layer of fresh material,
 see Figs. 10 and 12.
 GSGW moves under the levitating belt confined in meridional direction between the two horizontal lines.  
 GSGW rises pressure under its crest.
 The rise of pressure under crest $\Delta p,$
 normalized to hydrostatic pressure $p$ at the bottom of fuel, is $\Delta p/p\sim 10^{-4}.$
 This rise is significant
 since it lasts rather long $\Delta t\sim (\lambda/2)/v_{GSGW}\sim 10^{-2}-10^{-3}$s;
 here $\lambda\sim 60$ km is wavelength of GSGW
 and $v_{GSGW}\sim 10^8$ cm/s is its velocity (the red arrow) relative to matter corotating with neutron star.
 This significant and extended time rise of pressure and temperature
  ignites fuel in a point when fuel in the bottom is very near to the ignition state.     }
\end{figure}

 {\bf The importance of allowance for the stellar rotation and general relativity effects.}
 Observations of the millisecond pulsar/burster SAX J1808.4-3658 showed that this transient X-ray source
   in a binary system with a period of 2 h (Chakrabarty and Morgan 1998) occasionally exhibits X-ray bursts
     associated with nuclear explosions (in 't Zand et al. 1998), manifesting itself in a different state
      as an X-ray pulsar with a period of 2.49 ms (Wijnands and van der Klis 1998).
 During its activity, the source's luminosity changed rapidly from $6\times 10^{36}$
   to $2\times 10^{34}$ erg s$\!^{-1},$ with the luminosity having been halved
     in a time of the order of several hours (Gilfanov et al. 1998).
 This example clearly demonstrates the following:
  (a) the mechanism of bursters does not work all the time but only under certain conditions
   (possibly, when a giant solitary wave appears),
  (b) neutron stars rotate rapidly and their rotation velocity on the surface can reach 1/5--1/4
    of the Keplerian velocity,
  and (c) there is disk accretion onto the NS at a sufficiently high luminosity of the source --
    no other option is possible in such a close binary system.
 More than twenty neutron stars with weak magnetic fields and rapid rotation are known to date
  (van der Klis 2006).
 We consider the influence of stellar rotation on the structure
 of the levitating layer and its base in a different paper (Inogamov and Sunyaev 2011).
 We show that the stellar rotation velocity observed in these objects does not change radically
  the results of IS99 and this paper, but it should be taken into account.

 The dynamics of the spreading layer and its base is complex, while the physical conditions in it
  are very far from those in laboratory experiments and commonly considered problems
   related to the atmosphere and ocean as well as to the motion of aircraft and rockets
    and the braking of asteroids and satellites in the atmosphere. In this situation,
     it is natural to take the first steps by neglecting the effects of general and special relativity.

 Nevertheless, allowance should be made for the fact that the luminosity of the spreading layer
  is twice that of an extended accretion disk even for a non-rotating NS with a radius $3 R_g,$
  where $R_g =2 G M/c^2$ (Sunyaev and Shakura 1986). In the Newtonian problem, these two quantities are equal.
   The Keplerian velocity near the surface in the Schwarzschild metric is more than half the speed of light,
   while it is close to $0.4\,c$ in the Newtonian approach. The effects of rapid NS rotation
    (and counterrotation) on the luminosity of the spreading (or boundary) layer
     were studied in detail by Sibgatullin and Sunyaev (2000) in terms of general relativity.
      These effects are great.

 {\bf The luminous flux, radiation-dominated plasma in the levitation belt, and photon bubbles.}
 Here, we investigate the hydrodynamic stability of the levitating layer and the underlying base
   (see Section 10 and the Appendix).
 We show that the levitating layer is close to a neutral stability:
   there are neither radial convection nor gravity waves.

 In atmosphere and ocean physics, a neutrally stable medium is called subtile.
 Convection under the dominance of radiation pressure would lead to a partial spatial separation
   of radiation and matter and to the formation of buoyant photon bubbles.
 In the base, the ratio of the radiation and thermal pressures decreases rapidly as the density rises.
 Nevertheless, the radiation pressure accounts for an appreciable fraction of the plasma pressure
   as long as the matter remains rapidly rotating one and this is despite the dramatic rise in density.

 It is unlikely that it will be possible to maintain the balance of three forces
   (gravity, the centrifugal force, and the force associated with the radiation pressure gradient)
     necessary for levitation, first, at a level of a few thousandths of the weight and,
       second, everywhere with the sign that provides the pressing of matter to the star.
 In IS99, we constructed a stationary solution with the satisfaction of the condition for compensation
    at a level of a few thousandths along the entire base of the levitating belt from the equator
      to latitude $\theta_\star$ (see Fig. 1).
 However, analysis of the meridional equilibrium in the base (see section 1.10)
  leads to the conclusion that the meridional extent of the base is slightly less
    than that of the levitating belt $\theta_\star.$
 As a result, the scenario with the hurling of matter upward along the radius (like a wind)
  in the equatorial zone and with the sinking of this matter back onto the NS surface
   at latitudes $\sim \theta_\star$ should be realized.
 The wind lifting height on the equator is limited, because the luminous flux diverges in radial directions
  in a spherical geometry. The divergence of the luminous flux is disregarded in the IS99 solution,
   which is constructed in the approximation of a thin layer on the NS surface
    without allowance for the curvature of the NS surface.

 Such a flux outgoing on the equator and returning at latitudes $\sim \theta_\star,$ first,
  enhances the advection of radiation (see Fig. 2) and, second, makes this circulation random.
 According to the IS99 calculations, the meridional rotation period of such a large-scale random circulation loop
  is of the order of the rotation period of a matter particle in the levitating belt around the NS.
   Accordingly, the instantaneous luminosity of the levitating belt
    as a function of the azimuthal angle $\varphi$ contains a significant contribution
     from the first Fourier harmonic in $\varphi.$
 This can explain the appearance of the upper peak in the kilohertz variability of radiation
  from the boundary layer (see also Sections 1.8 and 1.11 below).
   Since the radiation circulation is random in nature, the Q factor of the peak turns out to be small.
    The question of why the lower peak appears is considered in
Section 1.11.

 {\bf Meridional force balance in the case of a thick base.}
  Consider the hydrostatic balancing of matter in the radial and meridional directions
   for a thick, rapidly rotating base.
 Rapid gas rotation in the case of a thick, massive base gives rise to a significant force of centrifugal rolling
  toward the equator. We compare two cases: (1) with a thick base and without any jump at the GSGW that reduces
 sharply the rotation velocity and (2) with a thin (in radius and mass) base in which there is a jump in
 velocity $v_\varphi$ (see Figs. 3 and 9 
  and the text below; in Figs. 3 and 9,  
   the jump is denoted by the letter J).
 Figures 9  
  and 3, 8,   
   respectively, illustrate that the thick base is several times thicker in radius and several
orders of magnitude thicker in surface density $\Sigma.$

It follows from Fig. 9  
 that, in the first case (thick base), the thickness of the layer of rapid (compared to
the NS rotation frequency) rotation is about 300 m.
 In the second case, this thickness is 100 m. The profiles in
Figs. 3, 8,  
 and 9  
  were obtained by integrating the equations derived in Sections 3--7. In particular, the
equations include the production of radiation energy, the outward radiation transport, and the dynamical
contribution of the light pressure. A base thickness of $\sim 300$ m means that an equatorial swelling about
300 m in height in the form of a belt rotating relative to the NS appears on the NS surface S
 (see Fig. 14).  

\begin{figure}[t]
 \includegraphics[width=1\columnwidth]{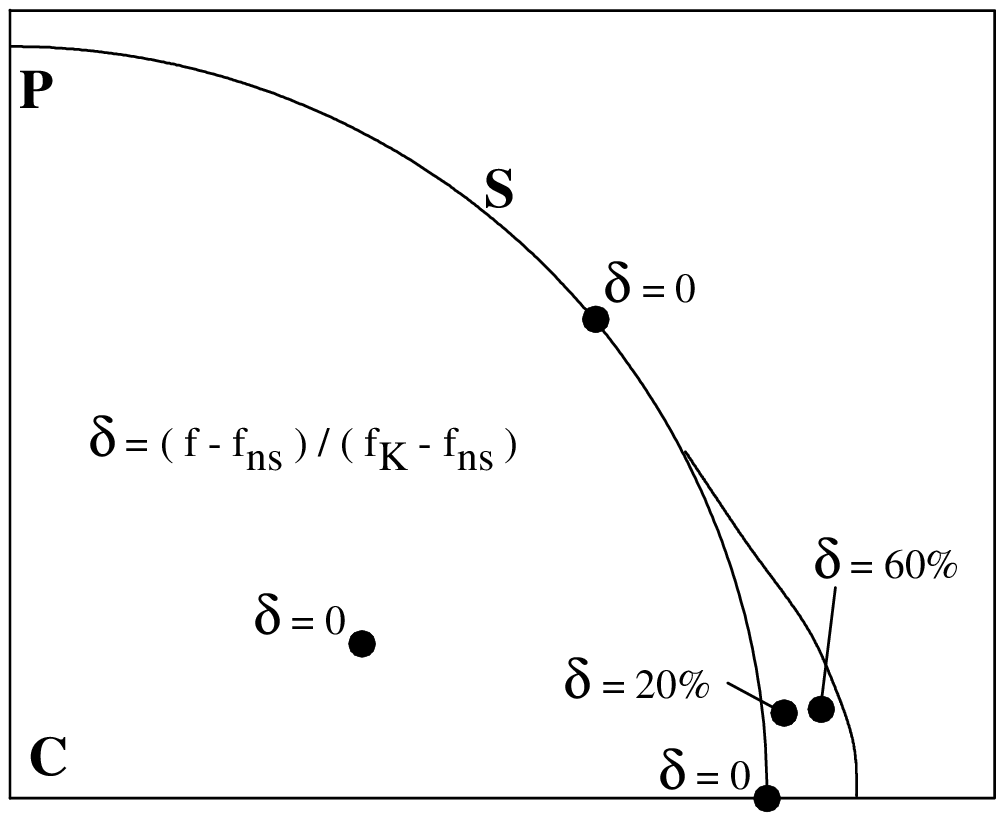}
 \caption{\label{fig:9-12}
  Formation of a very massive swelling on the NS
surface S due to nearly Keplerian rotation under the levitating layer L in the case of a spreading layer with a thick
base. In the swelling zone, the matter rotates differentially
and is supported by centrifugal forces.
      }
\end{figure}

The velocity components other than the rotation velocity (radial settling and meridional spreading) are low
and, hence, dynamically insignificant. When the centrifugal force is taken into account, the swelling is then
in hydrostatic equilibrium. The meridional scale of the swelling exceeds its radial extent. Accordingly, for
the description in the first approximation, it is admissible to use one-dimensional balances along the
vertical, as is done in Sections 3--6. The swelling in Fig. 14 
 is associated with high-density matter. This
qualitatively distinguishes it from the low-density levitating belt L in Fig. 1
 with a density $\rho_L \sim 10^{-3}$ g cm$\!^{-3}.$
As we see, not only the physical conditions on the horizon of helium flashes (the high-density horizon, see
below) but also the geometry of the zone with differential rotation change significantly in the case with a
thick base. The changes in geometry give rise to an equatorial swelling. Only high-density
 $(> 10^4$ g cm$\!^{-3})$ regions were attributed to this zone.

The radiative-meridional equilibrium in the second case (where there is a jump in velocity $v_\varphi,$ a thin
base) is discussed in Section 1.10. In this case, there is a slight rise of surface LA ``balancing on the verge
of a wind" above the NS surface. Surface LA separates the levitating matter and the base.

 \subsection{The Flow Structure: The Keplerian Flow--Star Interaction Mechanics}

We investigate the spreading layer that emerges in the region of contact between the accretion disk and the
surface of a nonmagnetic NS. The spreading layer consists of the upper rapidly rotating and lower slowly
rotating parts. The rotation frequency is of the order of the Keplerian one in the upper part and is
approximately equal to the stellar rotation frequency in the lower part. This layer is shown to be a complex
three-dimensional structure in the meridional $(\theta),$ radial $(r),$ and azimuthal $(\varphi)$ directions.
In the first approximation, the spreading layer is axisymmetric -- it forms an equatorial belt with azimuthal
rotation, meridional spreading, and radial accretion settling. In the next approximation, it is natural to
consider the large-scale azimuthal inhomogeneity. It is most likely this inhomogeneity that is responsible,
first, for the observed quasi-periodic oscillations at a frequency of the order of one kilohertz and, second,
for the splitting of these oscillations into two peaks -- upper and lower (see, e.g., van der Klis 2006). The
peaks shift in position within a fairly wide range, depending on the accretion rate $\dot M,$ with the
frequency difference equal to the NS rotation frequency retaining its value. The shifts of the peaks are
consistent with the views of a levitating belt whose width depends on $\dot M.$ In this case, the lower peak is
produced by the collisions of the inhomogeneity in the levitating belt with the disturbance associated with the
GSGW under the base (see Section 1.11 below).

The upper part of the spreading layer is broken up into two layers by surface LA. The upper layer above this
surface is extended along the radius through levitation (L) in the field of gravity. It was called the
levitating belt in IS99. This is a radiation-dominated layer in which levitation is achieved through an almost
complete radiation compensation for the sum of the gravitational and centrifugal accelerations (the Eddington
radiation support in the gravitational field). Therefore, surface LA may be called the boundary at which the
balancing of matter on the verge of a wind begins. Under this boundary, the radiation flux $q$ is slightly less
than its critical (i.e., local Eddington) value. Accordingly, as one recedes from LA downward, there is a rapid
rise in density with depth called in the paper condensation or accumulation (A, hence the designation LA).
Along the radius, the accumulation layer A is much thinner than the levitating belt L.

The accretion flow falls into the belt L through the neck that separates the disk and the spreading layer.
Above the neck, the angular momentum is transported over the disk to infinity by $\alpha$-friction. The angular
momentum $\approx \dot M \,v_K \,R$ brought by the flow into the spreading layer is transported downward to the
star; here, $v_K$ and $r$ are the Keplerian velocity and the stellar radius, respectively. This angular
momentum gives rise to a torque acting on the star. We analyze a complex system of gravity waves traveling in
dense underlying layers. They are excited dynamically under the action of tangential stress $w$ that is
produced during the distribution of torque per unit surface area of contact between the belt L and the star.
The existence of gravity waves is related to a sharp rise in density $\rho$ and to stratification stability. In
turn, the rise in $\rho$ and the stability are attributable to an extreme radiative heat transfer efficiency in
the optically thick case where the temperature of the spreading layer and high-density stellar envelopes is
much lower than the virial temperature under any heat releases up to the Eddington limit.

 \subsection{Observational Manifestations of the Spreading Layer}

The IS99 model of a levitating layer with bright bands of X-ray emission had a number of clear observational
predictions. The radiation from this layer should be highly variable on much shorter time scales than those for
the radiation from an extended accretion disk (for a discussion, see also Sunyaev and Revnivtsev 2000).

The bulk of the energy release in the spreading layer takes place at a great optical depth in its base. As a
result, the base forms a blackbody spectrum that subsequently passes through a hotter levitating plasma flow
with a finite optical depth for Thomson scattering and a negligible optical depth for bremsstrahlung processes.
Saturated Comptonization of photons from the base forms the observed radiation spectrum in the hotter
levitating layer similar to a diluted blackbody spectrum (Sunyaev and Titarchuk 1980). Detailed calculations of
the radiation spectrum forming in the spreading layer were performed by Grebenev and Sunyaev (2002) and
Suleimanov and Poutanen (2006). Gilfanov et al. (2003) and Revnivtsev and Gilfanov (2006) suggested and
successfully implemented a method for extracting the radiation spectrum of the boundary layer based on an
analysis of the rapidly varying radiation component in X-ray sources. Savolainen et al. (2009) and several
other observational groups are conducting a systematic search for manifestations of the boundary layer on the
surface of accreting neutron stars using INTEGRAL and RXTE data. A detailed theoretical study of the physical
processes in the base of the spreading layer must lead to new specific observational predictions.

Below, we make an attempt to solve the formulated problem. The picture of the flow as a whole is described in
Section 1. The results regarding the rapidly rotating low-density levitating belt L are summarized in Section
2. The dynamical and thermal equations are derived in Sections 3--7. Section 8 is devoted to the crisis of
friction with turbulent viscosity. Finally, in Sections 9--11, we analyze the influence of radiative viscosity
(Section 9), the generation of buoyancy-related gravity waves (Section 10), and supersonic effects (Section
11).



 Authors are very grateful to V. Astakhov for his careful translation.

\vspace{1cm}

\centerline{REFERENCES}

 1. N. Babkovskaya, A.~Brandenburg, and J.~Poutanen,
    Mon. Not. R. Astron. Soc. {\bf 386,} 1038 (2008).

 2. D. Chakrabarty and E. Morgan, Nature {\bf 394,} 346 (1998).

 3. R.L. Cooper and R. Narayan, Astrophys. J. {\bf 657,} L29 (2007).

 4. L.J. Dursi, A.C. Calder, A.~Alexakis, et al., arXiv:astro-ph/0207595 (2002).

 5. D.K. Galloway, M.P. Muno, J.M.~Hartman, et al., Astrophys. J. Suppl. Ser. {\bf 179,} 360 (2008).

 6. M. Gilfanov, M. Revnivtsev, R. Sunyaev, and E. Churazov, Astron. Astrophys. {\bf 338,} L83 (1998).

 7. M. Gilfanov, M. Revnivtsev, and S.~Molkov, Astron. Astrophys. {\bf 410,} 217 (2003).

 8. S.A. Grebenev and R.A. Sunyaev, Astron. Lett. {\bf 28,} 150 (2002).

 9. N.A. Inogamov and R.A. Sunyaev, Pis'ma Astron. Zh. {\bf 25,} 323 (1999) [Astron. Lett. {\bf 25,} 269 (1999)].

 10. N.A. Inogamov and R.A. Sunyaev, Pis'ma Astron. Zh. (2011) (in press).

 11. R. Kippenhahn and H.-C. Thomas, Astron. Astrophys. {\bf 63,} 265 (1978).

 12. M. van der Klis, in Compact Stellar X-ray Sources, Ed. by W.H.G. Lewin and M. van der Klis
     (Cambridge Univ., Cambridge, 2006), p. 39; arXiv:astro-ph/0410551.

 13. K.R. Lang, Astrophysical Formulae (Springer, Berlin, Heidelberg, New York, 1974).

 14. W.H.G. Lewin, J. van Paradijs, and R.E.~Taam, Space Sci. Rev. {\bf 62,} 223 (1993).

 15. M. Linares, A. Watts, D.~Altamirano, et al., Astrophys.J. Lett. {\bf 719,} L84 (2010).

 16. F. Peng, E.F. Brown, and J.W.~Truran, Astrophys. J. {\bf 654,} 1022 (2007).

 17. R. Popham and R.A. Sunyaev, Astrophys. J. {\bf 547,} 355 (2001).

 18. M.G. Revnivtsev and M.R.~Gilfanov, Astron. Astrophys. {\bf 453,} 253 (2006).

 19. R. Rosner, A. Alexakis, Y.-N.~Young, et al., arXiv:astro-ph/0110684 (2001).

 20. P. Savolainen, D.C. Hannikainen, O.~Vilhu, et al., Mon. Not. R. Astron. Soc. {\bf 393,} 569 (2009).

 21. H. Schlichting, Grenzschicht-Theorie (G.~Braun, Karlsruhe, 1965) [in German].

 22. N.I. Shakura and R.A. Sunyaev, Astron. Astrophys. {\bf 24,} 337 (1973).

 23. N.R. Sibgatullin and R.A. Sunyaev, Astron. Lett. {\bf 26,} 699 (2000).

 24. A. Spitkovsky, Y. Levin, and G.~Ushomirsky, Astrophys. J. {\bf 566,} 1018 (2002).

 25. L. Spitzer, Phys. of Fully Ionized Gases, 2nd rev. ed. (Intersci. Publ., New York, 1962).

 26. T. Strohmayer and L. Bildsten, in: Compact Stellar X-ray Sources, Ed. by W.H.G.~Lewin and M.~van der Klis
     (Cambridge Univ., Cambridge, 2006), p. 113.

 27. V. Suleimanov and J. Poutanen, Mon. Not. R. Astron. Soc. {\bf 369,} 2036 (2006).

 28. R. Sunyaev and M. Revnivtsev, Astron. Astrophys. {\bf 358,} 617 (2000).

 29. R.A. Sunyaev and L.G.~Titarchuk, Astron. Astrophys. {\bf 86,} 121 (1980).

 30. R.A. Sunyaev and N.I. Shakura, Sov. Astron. Lett. {\bf 12,} 117 (1986).

 31. J.L. Tassoul, Theory of Rotating Stars (Princeton Univ., Princeton, 1978; Mir, Moscow, 1982).

 32. S. Weinberg, Gravitation and Cosmology: Principles and Applications of the General Theory of Relativity
     (Wiley, New York, 1972).

 33. R. Wijnands and M.~van der Klis, Nature {\bf 394,} 344 (1998).

 34. J.J.M. in't Zand, J.~Heise, J.M.~Muller, et al., Astron. Astrophys. {\bf 331,} L25 (1998).

\end{document}